\documentclass[11pt,aps,superscriptaddress,preprintnumbers,amsmath,amssymb,nofootinbib]{revtex4}
\usepackage{epsfig}  
\usepackage{graphicx}
\usepackage{hyperref}

\hypersetup{
    colorlinks=true,
    citecolor=red,
    linkcolor=blue,
    filecolor=green,      
    urlcolor=magenta,
}
\usepackage{color}
\usepackage{float}
\usepackage{amsfonts}
\usepackage{amsmath}
\usepackage{slashed}
\usepackage{soul}
\usepackage{comment}
\def\beq{\begin{equation}}
\def\eeq{\end{equation}}
\def\bea{\begin{eqnarray}}
\def\eea{\end{eqnarray}}
\def\nn {\nonumber}
\newcommand{\bpl}{\beta_{+}}
\newcommand{\bmi}{\beta_{-}}
\newcommand{\mm}{m_{-}}
\newcommand{\mpl}{m_{+}}
\newcommand{\qsq}{q^{2}}
\def\re{{\rm Re}}  
\newcommand{\la}{\lambda}
\newcommand{\lone}{\ell_1}

\newcommand{\ltwo}{\ell_{2}}
\newcommand{\ltwobar}{\bar{\ell}_{2}}
\newcommand{\matel}[3]{\langle #1|#2|#3\rangle}
\large

\begin{document} 
\hfill{UdeM-GPP-TH-22-296}

\title{Lepton Flavor Violating $B\to K^*_2(1430)\mu^{\pm}\tau^{\mp}$ Decays}

\author{Suman Kumbhakar}
\email{suman.kumbhakar@umontreal.ca}
\affiliation{Physique des Particules, Université de Montréal, 1375 Ave.Thérèse-Lavoie-Roux, Montréal, QC H2V 0B3}

\author{Ria Sain}
\email{riasain@rnd.iitg.ac.in}
\affiliation{Department of Physics, Indian Institute of Technology Guwahati, Assam 781039, India}

\author{Juhi Vardani}
\email{vardani.1@iitj.ac.in}
\affiliation{Indian Institute of Technology Jodhpur, Jodhpur 342037, India}

\date{July 19, 2023}

\begin{abstract}

A number of measurements in decays induced by the semileptonic $b\to s$ and $b\to c$ transitions hint towards a possible role of new physics in both sectors.
 Motivated by these anomalies, we investigate the lepton flavor violating $B\to K^*_2 (1430)\mu^{\pm}\tau^{\mp}$ decays. We calculate the two-fold angular distribution of $B\to K^*_2\ell_1\ell_2$ decay in presence of vector, axial-vector, scalar and pseudo-scalar new physics interactions. We then compute the branching fraction and lepton forward-backward asymmetry in the framework of $U^{2/3}_1$ vector leptoquark which is a viable solution to the current $B$ anomalies. We find that the upper limits are $\mathcal{B}(B\to K^*_2\mu^-\tau^+)\leq 1.64\times 10^{-7}$ and $\mathcal{B}(B\to K^*_2\mu^+\tau^-)\leq 0.60\times 10^{-7}$ at $90\%$ C.L.
\end{abstract}
 
\maketitle 
\newpage
\section{Introduction} 
One of the most exciting experimental results in recent times in flavor physics is the violation of lepton flavor universality (LFU) in semileptonic $B$ meson decays (for a recent review, see Ref.~\cite{London:2021lfn}). These decays are mainly mediated by neutral current transition $b\to s\ell^+\ell^-$ ($\ell=e,\, \mu$) and charged current transition $b\to c\ell\bar{\nu}$ ($\ell=e,\, \mu,\, \tau$). However,  in $b\to s\ell^+\ell^-$ sector, the recent measured values of the LFU ratios $R_{K^{(*)}} = \mathcal{B}(B\to K^{(*)}\mu^+\mu^-)/\mathcal{B}(B\to K^{(*)}e^+e^-)$ by the LHCb collaboration~\cite{LHCb:2022zom,LHCb:2022qnv} are now consistent with their  Standard Model (SM) predictions. Still, there are other discrepancies  reported in the observables related to $b\to s\mu^+\mu^-$ decays such as the branching ratio of $B_s\to \phi\mu^+\mu^-$,  angular observables  in $B\to K^*\mu^+\mu^-$ decay~\cite{LHCb:2015wdu, LHCb:2021vsc, LHCb:2013ghj, Aaij:2020nrf}. These deviations are at the level of $\sim 2$-$4\sigma$. A number of studies has been performed to explain all these deviations in both model-independent (see for example Refs.~\cite{Alok:2019ufo,Altmannshofer:2021qrr,Carvunis:2021jga,Alguero:2021anc,Geng:2021nhg,Hurth:2021nsi,Ciuchini:2022wbq,SinghChundawat:2022ldm,SinghChundawat:2022zdf,Alguero:2023jeh}) as well as in specific new physics (NP) models such as $Z'$ boson~\cite{Crivellin:2015era,Chiang:2017hlj,Datta:2017ezo,Navarro:2021sfb,Alok:2022pjb,Lee:2022sic}, Leptoquark~\cite{Alok:2017sui,Becirevic:2017jtw,Fornal:2018dqn} etc.
    
    On the other hand, the flavor ratios in $b\to c\ell\bar{\nu}$ transition are defined as $R_{D^{(*)}} = \mathcal{B}(B\to D^{(*)}\tau\bar{\nu})/\mathcal{B}(B\to D^{(*)}\ell\bar{\nu})$, where $\ell = e$ or $\mu$~ \cite{Lees:2012xj,Lees:2013uzd,Huschle:2015rga,Sato:2016svk,Hirose:2016wfn,Abdesselam:2019dgh,Aaij:2015yra,Aaij:2017uff,Aaij:2017deq}. The present world averages of $R_{D}$ and $R_{D^*}$, including the correlation, are $\sim 3\sigma$ away from their SM predictions~\cite{HFLAV:2022pwe}. Besides, the measured value of the ratio $R_{J/\psi} = \mathcal{B}(B_c\to J/\psi\tau\bar{\nu})/\mathcal{B}(B_c\to J/\psi\mu\bar{\nu})$ is found to be $\sim 2\sigma$ higher than its SM value~\cite{Aaij:2017tyk}. Very recently, the LHCb has measured another ratio in baryonic decays $R_{\Lambda_c} = \mathcal{B}(\Lambda_b\to \Lambda_c\tau\bar{\nu})/\mathcal{B}(\Lambda_b\to \Lambda_c\mu\bar{\nu})$ which is consistent with its SM prediction~\cite{LHCb:2022piu}. In this context, various groups have proposed explanations in a model-independent approach~\cite{Hu:2018veh,Alok:2019uqc,Asadi:2019xrc,Murgui:2019czp,Bardhan:2019ljo,Blanke:2019qrx,Shi:2019gxi,Cheung:2020sbq,Iguro:2022yzr} and also in different NP models. Some of the simplified NP models may include a new vector boson~\cite{Greljo:2015mma,Boucenna:2016wpr,Matsuzaki:2017bpp,Asadi:2018wea}, a charged scalar~\cite{Crivellin:2012ye,Celis:2012dk,Crivellin:2015hha,Celis:2016azn,Ko:2017lzd,Iguro:2017ysu,Biswas:2018jun,Martinez:2018ynq} or a Leptoquark~\cite{Sakaki:2013bfa,Fajfer:2015ycq,Li:2016vvp,Barbieri:2016las,Calibbi:2017qbu}.

Several attempts are made to come up with a combined explanation of anomalies in both neutral current and charged current transitions~\cite{Bhattacharya:2016mcc,Alok:2017jaf,Kumar:2018kmr,Cornella:2019hct,Crivellin:2019dwb,Angelescu:2021lln,Belanger:2022kvj}. Among these, the Leptoquark (LQ) solution is quite interesting and popular. However, the LQ could be either a scalar or a vector depending on its spin. A review on all kinds of LQs can be found in Ref.~\cite{Dorsner:2016wpm}. Among all the LQs, the vector $U_1$ with charge +2/3 and a $SU(2)_L$ singlet is a very good candidate to explain all these deviations simultaneously.

The hints of LFU violation in $B$ decays automatically lead to a possibility of having lepton flavor violation (LFV) in $B$ decays~\cite{Glashow:2014iga,Guadagnoli:2015nra,Crivellin:2015era,Becirevic:2016zri,Das:2019omf,Bordone:2021usz}. In this work, we study the flavor violation in $B\to K^*_2\ell_1\ell_2$, where $K^*_2$ is higher excited spin-2 state of $K^*$ meson. The tensor meson $K^*_2$ has additional polarizations compared to the $K^*$ meson which could provide new kinematical quantities that are sensitive to the NP. Therefore, $B\to K^*_2$ decay could be examined as a complementary decay process to look for new helicity structures compared to the recently measured $B\to K^{*0}\mu\tau$ decay~\cite{LHCb:2022wrs}. Our main aim is to investigate the LFV properties of $B\to K^*_2\ell_1\ell_2$ decays if it could provide any interesting features. This decay is mediated by $b\to s\ell_1\ell_2$ where we consider $\ell_1$, $\ell_2 = \mu$ or $\tau$. The flavor conserving $B\to K^*_2\ell^+\ell^-$ decay has already been well studied within the SM as well as in the NP models~\cite{RaiChoudhury:2006bnu,Hatanaka:2009gb,Hatanaka:2010fpr,Li:2010ra,Lu:2011jm,Aliev:2011gc,Das:2018orb}. In particular, BaBar~\cite{BaBar:2003aji} and Belle~\cite{Belle:2002ekk} have already observed the radiative decay $B\to K^*_2\gamma$ and the measured branching ratio is as comparable to that of $B\to K^*\gamma$. In this paper, we calculate the two-fold differential decay distribution of $B\to K^*_2\ell_1\ell_2$ in the presence of vector, axial-vector, scalar and pseudo-scalar NP operators. Then we investigate the sensitivities of the branching ratio and the forward-backward asymmetry in the $U_1$ LQ model. In doing that, first we obtain the allowed NP parameter spaces of the model by using the most relevant constraints from the flavor observables. Finally, we compute the upper limits of these two observables in the $U_1$ model.

The paper is organized as follows. In sec.II, we start with the $b\to s\ell_1\ell_2$ NP Hamiltonian and calculate two-fold decay distribution of $B\to K^*_2\ell_1\ell_2$ decay. 
In sec.III, we discuss the $U_1$ LQ model and investigate its effect on the observables of $B\to K^*_2\mu^{\pm}\tau^{\mp}$ decays.
In sec. IV, we present our conclusions.  

\section{Decay distribution of $B\to K^*_2\ell_1\ell_2$}
The LFV $ b\to s\ell_1\ell_2 $ transition is not allowed in the SM. Therefore, the NP effective Hamiltonian for $ b\to s\ell_1\ell_2 $ transition can be written as
\begin{equation}
\mathcal{H}_{\rm eff} = - \frac{\alpha_{\rm em} G_F}{\sqrt{2} \pi} V_{ts}^* V_{tb} \sum_{i=V,A,S,P} (C_i\mathcal{O}_i +C^{\prime}_i\mathcal{O}^{\prime}_i) ,
\end{equation}
where $\alpha_{\rm em}$ is the fine-structure constant, $G_F$ is the Fermi constant, $V_{ts}$ and $V_{tb}$ are the Cabibbo-Kobayashi-Maskawa (CKM) matrix elements. The $\mathcal{H}_{\rm eff}$ contains four types of NP operators, vector(V), axial-vector(A), scalar(S) and pseudo-scalar(P), which can be expressed as 
\begin{eqnarray}\label{eq:operatorWC}
\mathcal{O}^{(\prime)}_V &=& [\bar{s}\gamma^{\mu}P_{L(R)}b] [\ell_2 \gamma_{\mu}\ell_1], \quad  \mathcal{O}^{(\prime)}_A = [\bar{s}\gamma^{\mu}P_{L(R)}b] [\ell_2 \gamma_{\mu}\gamma_5\ell_1], \nonumber \\ 
\mathcal{O}^{(\prime)}_S &=& [\bar{s}P_{R(L)}b] [\ell_2 \ell_1], \quad  \mathcal{O}^{(\prime)}_P = [\bar{s}P_{R(L)}b] [\ell_2 \gamma_5\ell_1]. 
\end{eqnarray} 
Here $P_{L(R)} = (1 \mp \gamma_5)/2$ are the chirality operators.
We note that the electromagnetic penguin operator $\mathcal{O}_7$ (appears in $b \to s \ell \ell$ decay):
\begin{equation}
    \mathcal{O}_{7} = \frac{m_b}{e}(\bar{s}\sigma_{\mu\nu}P_R b)F^{\mu\nu} 
\end{equation}
cannot generate LFV contribution due to the universality of electromagnetic interaction.
The NP Wilson coefficients (WCs) $C^{(\prime)}_{V,A,S,P}$ contain the short distance physics whereas the long-distance physics are embedded into the $B\to K^*_2$ hadronic matrix elements. For $B\to K^*_2$ transition, the hadronic matrix elements for V and A currents can be parameterized in terms of four form factors $V(q^2)$ and $A_{0,1,2}(q^2)$. These can be written as~\cite{Wang:2010ni}
 \begin{eqnarray}
  \langle K_2^*(k, \epsilon^*)|\bar s\gamma^{\mu}b|\overline B(p)\rangle
  &=&-\frac{2V(q^2)}{m_B+m_{K_2^*}}\epsilon^{\mu\nu\rho\sigma} \epsilon^*_{T\nu}  p_{\rho}k_{\sigma}, \nonumber\\
  \langle  K_2^*(k,\epsilon^*)|\bar s\gamma^{\mu}\gamma_5 b|\overline B(p)\rangle
   &=&2im_{K_2^*} A_0(q^2)\frac{\epsilon^*_{T } \cdot  q }{ q^2}q^{\mu} + i(m_B+m_{K_2^*})A_1(q^2)\left[ \epsilon^{*\mu}_{T}
    -\frac{\epsilon^*_{T } \cdot  q }{q^2}q^{\mu} \right] \nonumber\\
    &&-iA_2(q^2)\frac{\epsilon^*_{T} \cdot  q }{  m_B+m_{K_2^*} }
     \left[ (p+k)^{\mu}-\frac{m_B^2-m_{K_2^*}^2}{q^2}q^{\mu} \right],
\end{eqnarray}
where $p$ and $k$ are the four momentum of $B$ and $K_2^{*}$ mesons, respectively. However, applying equation of motion, it can be shown that the matrix element of $B\to K^*_2$ transition for the scalar interaction ($\bar{s}b$) vanishes and that for the pseudo-scalar interaction leads to 
\begin{equation}
 \langle K_2^*(k, \epsilon^*)|\bar s\gamma_{5}b|\overline B(p)\rangle =  -\frac{2i m_{K^*_2} A_0(q^2)}{m_b + m_s} (\epsilon^*_{T} \cdot  q).
\end{equation}
The polarization vector of $K^*_2$ is discussed in Appendix~\ref{appenb}. All form-factors have been calculated within the perturbative QCD approach~\cite{Wang:2010ni} and also in Light-Cone QCD sum rule technique~\cite{Cheng:2010hn,Wang:2010tz,Yang:2010qd,Aliev:2019ojc}. In this work, we follow the latest values of form factors derived using the light cone QCD sum rule (LCSR) from Ref~\cite{Aliev:2019ojc}. Within this method, each form factor can be expressed as follows 
\begin{equation}
\label{eqn:FormFactor}
F^{B\to T}(q^2)=\frac{1}{1-q^2/m^2_{R,F}}\sum_{n=0}^{1}\alpha_n^F\left[z(q^2)-z(0)\right]^n,
\end{equation}
where $z(s)=\frac{\sqrt{t_+-s}- \sqrt{t_+-t_0}}{\sqrt{t_+-s}+ \sqrt{t_+-t_0}}$, $t_{\pm}=(m_B\pm m_{K_2^*})^2$, $t_0=t_+(1-\sqrt{1-t_-/t_+})$ and $m_{R,F}$ is the resonance mass associated with the quantum numbers of corresponding form factor. The fit parameters $\alpha_n^F$ are given in Table~\ref{tab:FormFactor}.
\begin{table}[h!]
\begin{tabular}{|c|c|c|}
\hline
Form Factor & $\alpha_0$ & $\alpha_1$\\
\hline
$V^{B\to K_2^*}$ & $ 0.22^{+0.11}_{-0.08}$ &$-0.90^{+0.37}_{-0.50}$\\
\hline
$A_0^{B\to K_2^*}$ & $0.30^{+0.06}_{-0.05}$ & $-1.23^{+0.23}_{-0.23}$\\
\hline
$A_1^{B\to K_2^*}$ & $0.19^{+0.09}_{-0.07}$ & $-0.46^{+0.19}_{-0.25}$\\
\hline
$A_2^{B\to K_2^*}$ & $0.11^{+0.05}_{-0.06}$ & $-0.40^{+0.23}_{-0.16}$\\
\hline
\end{tabular}
\caption{Fit parameters for $B\to K_2^*$ form factors using LCSR approach.}
\label{tab:FormFactor}
\end{table}

The three body $B \to K_2^* \ell_1 \ell_2$ decay can be described by the leptonic polar angle $\theta_{\ell}$ and leptonic mass squared $q^2 = (p-k)^2$. We take the $\theta_{\ell}$ angle as the angle made by the $\ell_1$ lepton w.r.t. to the di-lepton rest frame. In terms of these two variables, we find the two-fold differential decay distribution as follows 
\begin{equation}
    \frac{d^2\Gamma}{dq^2 d \cos \theta_\ell}=A(q^2) + B(q^2) \cos \theta_\ell + C(q^2) \cos ^2\theta_\ell,
    \label{dist}
\end{equation}
where
\begin{eqnarray}
C(q^2)&=&\frac{3}{8}\bpl^2 \bmi^2  \left\lbrace\left(|A_L^{\parallel}|^2+|A_L^{\perp}|^2-2|A_L^{0}|^2\right)+\left(L\to R\right)\right\rbrace,\\\label{dist:C}
B(q^2)&=&\frac{3}{2} \bmi \bpl\left\lbrace \re\left[A_{L}^{\perp *} A_L^{\parallel} -(L \to R)\right]+ \frac{m_+m_-}{q^2} \re\left[A_{L}^{0*} A_L^{t} +(L \to R)\right]\right.\nn\\
&& \left. + \frac{m_{+}}{\sqrt{\qsq}} \re\left[A_S^* (A_L^0 +A_R^0) \right]-\frac{\mm} {\sqrt{\qsq}} \re\left[A_{SP}^* (A_L^0 -A_R^0) \right]\right\rbrace,\\
A(q^2) &=& \frac{3}{4}\left\lbrace\frac{1}{4}\left[\left(1+\frac{\mpl^2}{q^2}\right)\bmi^2 +\left(1+\frac{\mm^2}{q^2}\right)\bpl^2\right]\left(|A_L^{\parallel}|^2+|A_L^{\perp}|^2+(L\to R)\right)\right.\nn\\
&&+\frac{1}{2}\left(\bpl^2 +\bmi^2\right)\left(|A_L^{0}|^2+|A_R^{0}|^2\right)\nn\\
&&+\frac{4m_1 m_2}{q^2} \re\left[ A_R^0 A_L^{0*} +A_R^{\parallel} A_L^{\parallel *}+A_R^{\perp} A_L^{\perp*}-A_L^t A_R^{t*}\right]\nn\\
&&+\frac{1}{2}\left(\bmi^2+\bpl^2-2\bmi^2\bpl^2\right)\left(|A_L^t|^2 + |A_R^t|^2\right) +\frac{1}{2} \left(|A_{SP}|^2 \bmi^2 +|A_{S}|^2 \bpl^2\right)\nn\\
&&\left.+\frac{2\mm}{ \sqrt{q^2}}\bpl^2 \re\left[A_S(A_L^t+A_R^t)^*\right]- \frac{2\mpl}{\sqrt{q^2}} \bmi^2 \re\left[A_{SP}(A_L^t-A_R^t)^*\right]\right\rbrace\label{dist:A}.
 \end{eqnarray}
Here $m_{\pm}=(m_1 \pm m_2)$, $\beta_{\pm} = \sqrt{1-\frac{(m_{\ell_{1}} \pm m_{\ell_{2}})^2}{q^2}} $ and the expressions of transversity amplitudes $A$'s are given in Appendix~\ref{Tamp}. In Appendix~\ref{LepHel}, we describe the lepton helicity amplitudes used in our analysis. Integrating Eq.~(\ref{dist}) over $\theta_\ell$, we get the differential decay rate
\begin{equation}
    \frac{d\Gamma}{d q^2}= 2\left(A  + \frac{C}{3}\right)
\end{equation}
whereas the lepton forward-backward asymmetry is found to be
\begin{equation}
A_{\rm FB}(q^2)= \frac{1}{d\Gamma/dq^2}\left(\int_0^1 d\cos\theta_\ell\frac{d^2\Gamma}{d\cos\theta_\ell d q^2}-\int_{-1}^0 d\cos\theta_\ell \frac{d^2\Gamma}{d\cos\theta_\ell d q^2 }\right) = \frac{B}{2\left(A+\frac{C}{3}\right)}.
\end{equation}

\section{The Vector $SU(2)_L$ singlet LQ}
We consider the weak singlet vector LQ $U^{2/3}_1$ since it can give rise to a good explanation of both types of anomalies in $b\to c\ell\bar{\nu}$ and $b\to s\mu^+\mu^-$ transitions, simultaneously~\cite{Kumar:2018kmr,Cornella:2019hct,Angelescu:2021lln}. In fact, this LQ model attracts a lot of attention in the literature. The interaction Lagrangian between the $U_1$ LQ and the SM fermions can be written as~\cite{Dorsner:2016wpm}
\begin{equation}
\mathcal{L}_{U^{2/3}_1} = h^{ij}_L \bar{Q}_{iL}\gamma_{\mu}L_{jL}U^{\mu}_{1} +h^{ij}_R \bar{d}_{iR} \gamma_{\mu}\ell_{jR}U^{\mu}_1 + {\rm h.c.},
\end{equation}
where $h^{ij}_{L,R}$ are the LQ couplings with the generation indexes $i,j$, $Q_{L}$($L_{L}$) is the SM left-handed quark (lepton) doublet and $d_R$($\ell_R$) is the right-handed down quark (lepton) singlet. We consider the minimal framework where $h^{ij}_{R} = 0$. We also assume the LQ couplings are real and couplings to the first generation are zero. This leads to the following flavor structure of the LQ couplings
\begin{equation}
h_L = 
    \begin{pmatrix}
0 & 0 & 0\\
0 & h^{22}_L & h^{23}_L\\
0 & h^{32}_L  &  h^{33}_L 
\end{pmatrix}.
\end{equation}
To calculate the NP parameters $h_{L}$'s, one needs to consider all the relevant constraints from low energy observables. In $U_1$ LQ model, these constraints would include from $2q2\ell$ processes which involve second and third generations. In our analysis, we consider only those processes which contribute at the tree level. These observables can be divided into following categories
\begin{itemize}
    \item $b\to s\mu^+\mu^-$ data : In presence of $U_1$ LQ, the vector and axial-vector NP WCs of $b\to s\mu^+\mu^-$ can be written as
\begin{equation}
C^{\mu\mu}_V = -C^{\mu\mu}_{A} = \frac{\pi}{\sqrt{2}G_F V_{tb} V^*_{ts}\alpha_{em}}\frac{h^{22}_L\,h^{32*}_L}{M^2_{U_1}},
\end{equation}
where $C^{\mu\mu}_{V(A)}$ denotes the NP WC of $[\bar{s}\gamma^{\alpha}P_{L}b] [\bar{\mu} \gamma_{\alpha}(\gamma_5)\mu]$ operator. From recent global fits of all $b\to s\mu^+\mu^-$ data, the most preferred value is  $C^{\mu\mu}_V =-C^{\mu\mu}_{A}=-0.19\pm 0.06$ \cite{Alguero:2023jeh} which is used in our calculation.

\item $b\to c(u)\ell\bar{\nu}$ data: In case of $b\to c\tau\bar{\nu}$, the NP WC can also be expressed in following form
\begin{equation}
C^{\tau}_{V_L} = \frac{1}{2\sqrt{2}G_F M^2_{U_1}} h^{33*}_L \left[h^{33}_L + (V_{cs}/V_{cb})h^{23}_L\right],
\end{equation}
where $C^{\tau}_{V_L}$ is the NP WC of the NP operator $\mathcal{O}_{V_L} = (\bar{c}\gamma_{\mu}P_Lb)(\bar{\tau}\gamma^{\mu}P_L\nu)$. In this sector, we consider four flavor ratios $R_D$, $R_{D^*}$, $R_{J/\psi}$ and $R_{\Lambda_c}$. The expressions for these ratios can be found in Ref.~\cite{Alok:2017qsi,Garcia-Duque:2022tti}. In addition, $\mathcal{B}(B_c\to \tau\bar{\nu})$ puts an important constraint on the NP parameter space, particularly for scalar and pseudo-scalar interactions. This decay has not yet been measured. However, the LEP data at Z peak imposes a constraint $\mathcal{B}(B_c\to \tau\bar{\nu}) < 10\%$~\cite{Akeroyd:2017mhr} and a $30\%$ limit can be achieved using the $B_c$ lifetime~\cite{Alonso:2016oyd}. In our analysis we use the $30\%$ bound on $B_c\to \tau\bar{\nu}$.
There are two more LFU ratios $R^{\mu/e}_D = \mathcal{B}(B\to D \mu\bar{\nu})/\mathcal{B}(B\to D e\bar{\nu})$ and $R^{e/\mu}_{D^*} = \mathcal{B}(B\to D^* e\bar{\nu})/\mathcal{B}(B\to D^* \mu\bar{\nu})$ have been measured~\cite{Belle:2015pkj,Belle:2017rcc}. These measurements are well in agreement with their SM predictions. In this case, the $U_1$ LQ contribute to $b\to c\mu\bar{\nu}$ decay and the NP WC looks as
\begin{equation}
C^{\mu}_{V_L} = \frac{1}{2\sqrt{2}G_F M^2_{U_1}} h^{32*}_L \left[h^{32}_L + (V_{cs}/V_{cb})h^{22}_L\right].
\end{equation}
The $U_1$ LQ can also contribute to charged current transition $b\to u\tau\bar{\nu}$. Here one potential observable is $\mathcal{B}(B\to \tau\bar{\nu})$. The $U_1$ model contribute the this decay through the following NP WC
\begin{equation}
C^{u}_{V_L} = \frac{1}{2\sqrt{2}G_F M^2_{U_1}} h^{33*}_L \left[h^{33}_L + (V_{us}/V_{ub})h^{23}_L\right],
\end{equation}
where $V_{ub}$ and $V_{us}$ are the CKM matrix elements. The measured value of the branching ratio is in agreement with its SM prediction $\mathcal{B}(B\to \tau\bar{\nu})|_{\rm SM} = (0.812\pm 0.054)\times 10^{-4}$~\cite{UTfit}. The current measured values of all these quantities are listed in Table~\ref{expt}.
\item LFV decays: The LFV decays of $B$ meson are mediated by $b\to s\tau^{\pm}\mu^{\mp}$ which occur at tree level under exchange of $U_1$ LQ. In this model, the NP WCs take the following for
\begin{equation}
C^{ij}_V = -C^{ij}_{A} = \frac{\pi}{\sqrt{2}G_F V_{tb} V^*_{ts}\alpha_{em}}\frac{h^{2i}_L\,h^{3j*}_L}{M^2_{U_1}}.
\end{equation}
We note that the $b\to s\mu^-\tau^+$ ($b\to s\mu^+\tau^-$) decay depends on $h^{23}_Lh^{32}_L$ ($h^{22}_Lh^{33}_L$) combination. This includes the current upper limits on $B\to K\mu^{\pm}\tau^{\mp}$, $B\to K^{*0}\mu^{\pm}\tau^{\mp}$ and $B_s\to \mu^{\pm}\tau^{\mp}$, listed in Table~\ref{expt}. In addition, there could be constraints from $\Upsilon(nS)\to \tau^{\pm}\mu^{\mp}$ and $\tau \to\mu\phi$ which are governed by  $b\bar{b} \to \mu^{\pm}\tau^{\mp}$ and $\tau \to \mu\bar{s}s$ respectively. For these decays, the NP WCs in the $U_1$ model can be written as
\begin{eqnarray}
C^{bb\mu\tau}_V &=& -C^{bb\mu\tau}_{A} = \frac{\pi}{\sqrt{2}G_F V_{tb} V^*_{ts}\alpha_{em}}\frac{h^{33}_L\,h^{32*}_L}{M^2_{U_1}},\nonumber\\
C^{ss\mu\tau}_V &=& -C^{ss\mu\tau}_{A} = \frac{\pi}{\sqrt{2}G_F V_{tb} V^*_{ts}\alpha_{em}}\frac{h^{23}_L\,h^{22*}_L}{M^2_{U_1}}.
\end{eqnarray}
Among three LFV decays of $\Upsilon(nS)$, the $\Upsilon(3S)$ decay puts the most strongest constraint which is used in our analysis. The expressions of all these LFV observables can be found in Ref.~\cite{Kumar:2018kmr}.
\item $b\to s\tau^+\tau^-$ decay: There could be important constraints from the rare decays $B\to K\tau^+\tau^-$ and $B_s \to \tau^+\tau^-$ which are mediated by $b\to s\tau^+\tau^-$ transition. In the $U_1$ LQ model, the NP WCs for this transition can be written as
\begin{equation}
C^{\tau\tau}_V = -C^{\tau\tau}_{A} = \frac{\pi}{\sqrt{2}G_F V_{tb} V^*_{ts}\alpha_{em}}\frac{h^{33}_L\,h^{23*}_L}{M^2_{U_1}}.
\end{equation}
The expressions for the branching fractions can be found in Ref.~\cite{Cornella:2019hct}. These decays are not observed so far, but we have the experimental upper limits, listed in Table~\ref{expt}.
\end{itemize}

\begin{table}[h!]
\begin{tabular}{|c|c|c|}
\hline
Sector & Observable & Measurement\\
\hline
 $b\to s\mu^+\mu^-$ & All $b\to s\mu^+\mu^-$  & $C^{\mu\mu}_9 = -C^{\mu\mu}_{10} = -0.19\pm 0.06$~\cite{Alguero:2023jeh}\\
\hline
&$R_D$ & $0.358\pm 0.025\pm 0.012$~\cite{HFLAV:2022pwe}\\
$b\to c(u)\ell\bar{\nu}$ &$R_{D^*}$ & $0.285\pm 0.010\pm 0.008$~\cite{HFLAV:2022pwe}\\
&$R_{J/\psi}$ & $0.71\pm 0.17\pm 0.18$~\cite{Aaij:2017tyk}\\
& $R_{\Lambda_c}$ & $0.242 \pm 0.026 \pm 0.040 \pm 0.059$~\cite{LHCb:2022piu}\\
& $R^{\mu/e}_D$ & $0.995 \pm 0.022 \pm  0.039$~\cite{Belle:2015pkj}\\
& $R^{e/\mu}_{D^*}$ & $1.04 \pm 0.05 \pm  0.01$~\cite{Belle:2017rcc}\\
& $\mathcal{B}(B\to \tau\bar{\nu})$ & $(1.09\pm 0.24)\times 10^{-4}$~\cite{Workman:2022ynf}\\
\hline
&$\mathcal{B}(B^+\to K^+\tau^-\mu^+)$ & $(0.8^{+1.9}_{-1.4}) \times 10^{-5}$; $< 4.5\times 10^{-5}$~\cite{BaBar:2012azg}\\
LFV &$\mathcal{B}(B^+\to K^+\tau^+\mu^-)$ & $(-0.4^{+1.4}_{-0.9})\times 10^{-5}$; $< 2.8\times 10^{-5}$~\cite{BaBar:2012azg}\\
&$\mathcal{B}(B^+\to K^{*0}\tau^+\mu^-)$ & $< 1.0\times 10^{-5}$~\cite{LHCb:2022wrs}\\
&$\mathcal{B}(B^{0}\to K^{*0}\tau^-\mu^+)$ & $< 8.2\times 10^{-6}$~\cite{LHCb:2022wrs}\\
&$\mathcal{B}(\Upsilon(3S)\to \tau^{\pm}\mu^{\mp})$ & $(-0.8^{+1.5 +1.4}_{-1.5-1.3})\times 10^{-6}$; $< 3.1\times 10^{-6}$~\cite{BaBar:2010vxb}\\
& $\mathcal{B}(\tau\to \mu\phi)$ & $< 8.4\times 10^{-8}$~\cite{Workman:2022ynf}\\
& $\mathcal{B}(B_s\to\tau^{\pm}\mu^{\mp})$ & $<3.4\times 10^{-5}$~\cite{LHCb:2019ujz}\\
 \hline
 $b\to s\tau^+\tau^-$ & $\mathcal{B}(B^+\to K^+\tau^+\tau^-)$ & $(1.31^{+0.66+0.35}_{-0.61-0.25}) \times 10^{-3}$; $<2.25\times 10^{-3}$~\cite{BaBar:2016wgb}\\
 & $\mathcal{B}(B_s\to \tau^+\tau^-)$ & $<5.2\times 10^{-3}$~\cite{Workman:2022ynf}\\
 \hline
\end{tabular}
\caption{List of observables and their measured values used into the $U^{2/3}_1$ LQ fit. All the upper limit are given at $90\%$ C.L.}
\label{expt}
\end{table}

\begin{figure}[h!]
\begin{tabular}{cc}
\includegraphics[width=85mm]{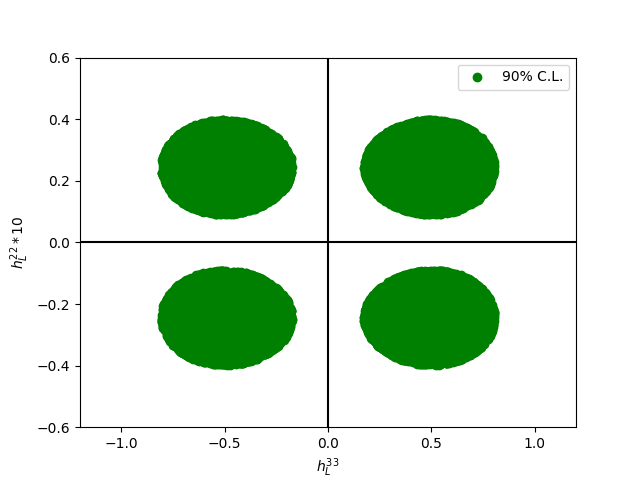} & \includegraphics[width=85mm]{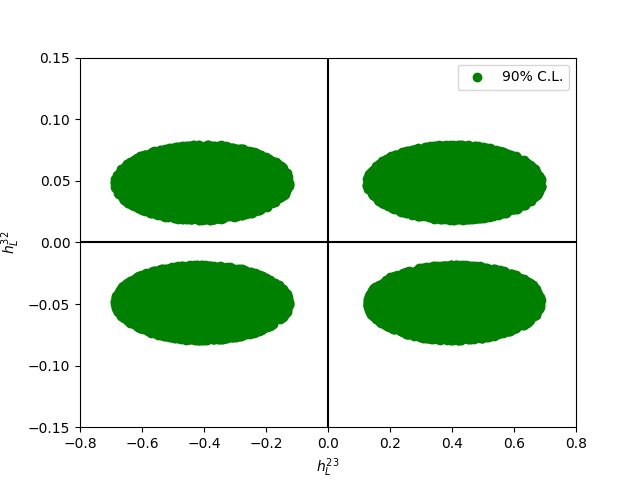}\\
\end{tabular}
\caption{The plot on the left (right) panel represents the allowed $90\%$ C.L. region in the $h^{22}_L$-$h^{33}_L$ ($h^{23}_L$-$h^{32}_L$) plane for $M_{U_1} =1.8$ TeV.}
\label{fig1}
\end{figure}

In our analysis, we assume the LQ mass to be $1.8$ TeV. This is the lower limit on the mass of vector LQ obtained from an analysis of LHC direct and indirect high $p_T$ searches~\cite{Angelescu:2021lln}. Now we perform a $\chi^2$ analysis to find the NP parameter space allowed by the current flavor data. The $\chi^2$ function is defined as 
\begin{equation}
\chi^2(h^{ij}_L) = \sum_k \frac{(\mathcal{O}^{\rm theory}_k (h^{ij}_L) - \mathcal{O}^{\rm expt}_k)^2}{\sigma^2_{\rm total, k}},
\end{equation}
where $\mathcal{O}^{\rm theory}_k$ is the theoretical prediction, $\mathcal{O}^{\rm expt}_k$ is the experimental central value and $\sigma_{\rm total, k}$ is the total uncertainty (theoretical and experimental errors are added in quadrature) for each observable. The measured values and the upper limits of each observable are listed in Table~\ref{expt} which went into the fit. However, we note that there are only a $90\%$ CL upper limit on the branching ratios of $B\to K^{*0}\tau^{\pm}\mu^{\mp}$, $\tau\to \mu\phi$, $B_s\to \tau^{\pm}\mu^{\mp}$ and $B_s\to \tau^+\tau^-$. In order to incorporate these observables into our fit, we take the branching ratio to be $(0.0\pm \rm {U.L.}/1.645)$.  The allowed NP couplings are obtained by minimizing the $\chi^2$ function. For the minimization, we use the CERN {\tt Minuit} library~\cite{James:1975dr}. We obtain the NP parameter spaces in the planes of $h^{22}_L$-$h^{33}_L$ and $h^{23}_L$-$h^{32}_L$  allowed at $90\%$ C.L. around the $\chi^2_{\rm min}$. These regions are shown in Fig.~\ref{fig1}. In particular, we note that the regions are symmetric for each plot.

\begin{figure}[h!]
\begin{tabular}{cc}
\includegraphics[width=80mm]{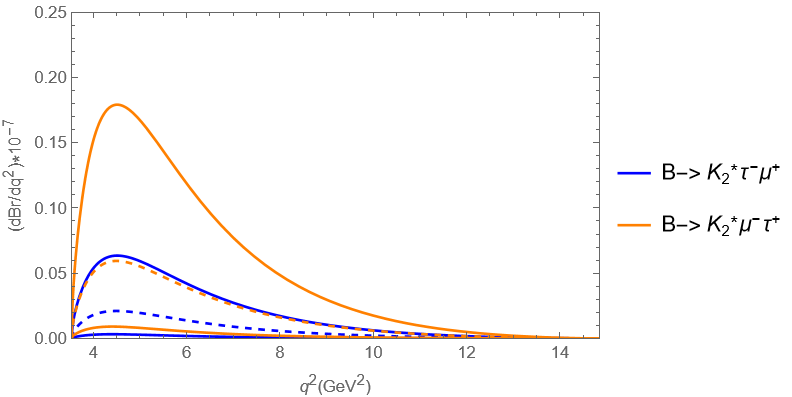} & \includegraphics[width=80mm]{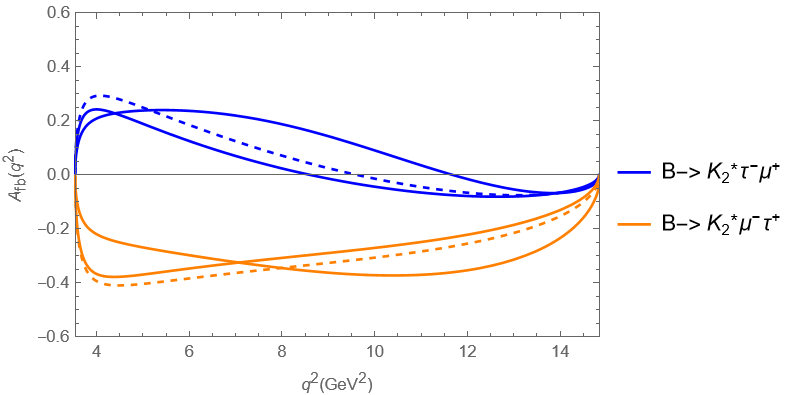}\\
\end{tabular}
\caption{The plots on the left and right panels represent the differential branching ratio and the forward-backward asymmetry as a function of $q^2$, respectively. These are the predictions for the benchmark points $h^{23}_Lh^{32}_L = 0.042$ and $h^{22}_Lh^{33}_L = 0.025$ which could give rise to maximum values of two observables at $90\%$ C.L}
\label{fig2}
\end{figure}

Now we calculate the maximum values of branching ratios and forward-backward asymmetries of $B\to K^*_2\mu^{\pm}\tau^{\mp}$ decays for the allowed NP parameter space. In order to do that, we scan over the allowed regions and find the benchmark point from each of the plots. From the left plot of Fig.~\ref{fig1}, we find $h^{22}_Lh^{33}_L = 0.025$ whereas from the right plot we get $h^{23}_Lh^{32}_L = 0.042$. At these values, we find the maximum values of branching fractions to be 
\begin{equation}
   \mathcal{B}(B\to K^*_2\mu^-\tau^+) \leq 1.64\times 10^{-7}, \quad   \mathcal{B}(B\to K^*_2\mu^+\tau^-) \leq 0.60\times 10^{-7}
   \label{brs}
\end{equation}
and those for forward-backward asymmetries are
\begin{equation}
   A_{\rm FB}(B\to K^*_2\mu^-\tau^+) \leq -0.347, \quad   A_{\rm FB}(B\to K^*_2\mu^+\tau^-) \leq 0.079.
\end{equation}
We also plot the differential branching ratio and the forward-backward asymmetry as a function of $q^2$ for the same NP benchmark points. The plots are shown in Fig.~\ref{fig2}. In these plots, the dashed curves represent the central values and solid curves indicates the widths of errors, mainly, due to the form factors. From both the differential branching ratio plot and Eq.~(\ref{brs}), it is evident that the order of magnitude is same for both decay. However the maximum value of $\mathcal{B}(B\to K^*_2\mu^+\tau^-)$ is less than  $\mathcal{B}(B\to K^*_2\mu^-\tau^+)$. From the $A_{\rm FB}$ plot, we find that
 the $A_{\rm FB}(q^2)$ of $B\to K^*_2\mu^-\tau^+$ is negative through out the whole $q^2$ region. However, the $A_{\rm FB}(q^2)$ of $B\to K^*_2\mu^+\tau^-$ has a zero-crossing point at $q^2 = 10$ GeV$^2$.

\section{Conclusions}
Inspired by the current $B$ physics anomalies, we investigate the LFV $B\to K^*_2\mu^{\pm}\tau^{\mp}$ decays. These decays are forbidden in the SM and hence any hint of these decays would imply a smoking gun signal of physics beyond the SM. In this work, we compute the two-fold angular distribution of $B\to K^*_2\ell_1\ell_2$ in presence of vector, axial-vector, scalar and pseudo-scalar NP interactions. From the differential decay distribution, we extract two observables: the branching fraction and the lepton forward-backward asymmetry. Finally, we calculate these observables in the vector $U_1$ LQ model which is a very popular model in explaining all the $B$ anomalies. In particular, we compute the upper limits of the branching fraction and the forward-backward asymmetry at $90\%$ confidence level. We find that the upper limits of branching fractions are $ \mathcal{O}(10^{-7})$. We hope these upper limits may be measurable at LHCb. 

\section*{Acknowledgments}
We thank Ashutosh Kumar Alok for useful discussions. This work was financially supported by NSERC of Canada (SK).
RS would like to acknowledge SERB sponsored project titled ``Probing New Physics Interactions" (CRG/2018/004889)  where this project started. RS would also acknowledge SERB National Postdoctoral Fellowship (NPDF) project PDF/2021/003328 for the support to complete the project.

\appendix

\section{Polarization of $K^*_2$}
\label{appenb}
The polarization $\epsilon^{\mu\nu}(n)$ of tensor meson $K_2^\ast$, which has four momentum $(k_0, 0, 0, \vec{k})$, can be written in terms of the spin-1 polarization vectors~\cite{Berger:2000wt}
\begin{eqnarray}
 \epsilon_{\mu\nu}(\pm 2) &=& \epsilon_\mu(\pm 1)\epsilon_\nu(\pm 1),\nn\\
 \epsilon_{\mu\nu}(\pm 1) &=& \frac{1}{\sqrt{2}}\left[\epsilon_\nu(\pm)\epsilon_\nu(0) + \epsilon_\nu(\pm)\epsilon_\mu(0) \right],\nn\\
 \epsilon_{\mu\nu}(0) &=& \frac{1}{\sqrt{6}}\left[\epsilon_\mu(+) \epsilon_\nu(-) + \epsilon_\nu(+) \epsilon_\mu(-) \right]+ \sqrt{\frac{2}{3}}\epsilon_\mu(0)\epsilon_\nu(0) ,
 \end{eqnarray} 
where the spin-1 polarization vectors are defined as
\begin{equation}
\epsilon_\mu(0) = \frac{1}{m_{K_2^\ast}}\left(\vec{k}_z,0,0,k_0\right)\, ,\quad \epsilon_\mu(\pm) = \frac{1}{\sqrt{2}}\left(0,1,\pm i, 0\right)\
\end{equation}

We study $B\to K^*_2\ell_1\ell_2$ decay where we have two leptons in the final state so in this case, the $n=\pm 2$ helicity states of the $K_2^\ast$
is not realized. Therefore a new polarization vector is introduced~\cite{Wang:2010tz}
\begin{equation}
\epsilon_{T\mu}(h) = \frac{\epsilon_{\mu\nu}p^\nu}{m_B}\, 
\end{equation}
 The explicit expressions of polarization vectors are
\begin{eqnarray}
\epsilon_{T\mu}(\pm 1) &=& \frac{1}{m_B}\frac{1}{\sqrt{2}}\epsilon(0).p  \epsilon_\mu(\pm) = \frac{\sqrt{\lambda}}{\sqrt{8}m_B m_{K^*_2}} \epsilon_\mu(\pm), \\
\epsilon_{T\mu}(0) &=& \frac{1}{m_B}\sqrt{\frac{2}{3}}\epsilon(0).p \epsilon_\mu(0) = \frac{\sqrt{\lambda}}{\sqrt{6}m_B m_{K^*_2}} \epsilon_\mu(0),
\end{eqnarray}
where $\lambda(m^2_B,m^2_{K^*_2},q^2) = m^4_B + m^4_{K^*_2} + q^4 -2(m^2_B m^2_{K^*_2}+m^2_Bq^2+m^2_{K^*_2}q^2)$ is the usual Kallen function.
On the other hand, the virtual gauge boson can have three types of polarization states, longitudinal, transverse and time-like, which have following components
\begin{equation}
\epsilon^\mu_V(0) = \frac{1}{\sqrt{q^2}}(-|\vec{q_z}|,0,0,-q_0)\, ,\quad \epsilon^\mu_V(\pm) = \frac{1}{\sqrt{2}}(0,1,\pm i, 0)\ ,\quad
\epsilon^\mu_V(t) = \frac{1}{\sqrt{q^2}}(q_0,0,0,q_z)\ 
\end{equation}
where $q^\mu=(q_0,0,0,q_z)$ is four momentum of gauge boson.

\section{Transversity Amplitudes}
\label{Tamp}
The vector and axial-vector transversity amplitudes can be expressed as
 \begin{eqnarray}
A_{0L,R} &=& N  \frac{\sqrt{\lambda}}{\sqrt6 m_Bm_{K_2^*}}\frac{1}{2m_{K^*_2}\sqrt {q^2}}\left[ (C_{V-}\mp C_{A-})
\left[(m_B^2-m_{K^*_2}^2-q^2)(m_B+m_{K^*_2})A_1 -\frac{\lambda}{m_B+m_{K^*_2}}A_2 \right] \right], \nonumber\\
A_{\perp L,R} &=& -\sqrt{2} N \frac{\sqrt{\lambda}}{\sqrt8m_Bm_{K_2^*}}\left[(C_{V+}\mp C_{A+})
 \frac{\sqrt \lambda V}{m_B+m_{K^*_2}}\right],   \nonumber \\
A_{\parallel L,R} &=& \sqrt{2} N\frac{\sqrt{\lambda}}{\sqrt{8} m_B m_{K_2^*}} \left[(C_{V-}\mp C_{A-}) (m_B+m_{K^*_2}) A_1\right], \nonumber \\
A_{Lt} &=&N\frac{\sqrt{\lambda}}{\sqrt{q^2}\sqrt{6}m_B m_{K^\ast_2}}\left[ \sqrt{\lambda}(C_{V-}-C_{A-}) A_0\right],\nonumber \\
A_{Rt} &=&N\frac{\sqrt{\lambda}}{\sqrt{q^2}\sqrt{6}m_B m_{K^\ast_2}}\left[ \sqrt{\lambda}(C_{V-}+C_{A-}) A_0\right],
\end{eqnarray}
where $C_{V\pm} = (C_{V}  \pm C_{V}^\prime)$, and $C_{A\pm} = (C_{A} \pm C_{A}^\prime)$.
The transversity amplitudes for scalar, pseudoscalar  interactions can be written as 
\begin{eqnarray}
  A_S  &=& 2 N \frac{\sqrt{\lambda}}{\sqrt{6}m_B m_{K^\ast_2}}\left [\sqrt{\lambda}(C_{S} - C_{S'}) A_0 \right],\nonumber \\
  A_{SP} &=& 2 N\frac{\sqrt{\lambda}}{\sqrt{6}m_B m_{K^\ast_2}}\left[\sqrt{\lambda}(C_P-C_{P'})A_0\right].
\end{eqnarray}
The normalization constant $N$ is given by
\begin{equation}
    N = \left[ \frac{G_F^2\alpha_e^2}{3\cdot 2^{10}\pi^5 m_B^3}|V_{tb}V_{ts}^\ast|^2 q^2\bpl \bmi \lambda(m^2_B,m^2_{K^*_2},q^2)^{1/2} \mathcal{B}(K_2^\ast\to K\pi) \right]^\frac{1}{2}.
\end{equation}


\section{Lepton helicity amplitudes \label{LepHel}}
Apart from the hadronic matrix elements, we also need the leptonic matrix elements to compute the decay distribution. To calculate these, we use the method outlined in refs.~\cite{Haber:1994pe, Gratrex:2015hna}. The leptonic matrix elements are defined as $\matel{\lone(\la_1) \ltwobar(\la_2)}{ \bar \ell   \; \Gamma^X \ell}{0}  = 
  \bar u(\la_1) \Gamma^X v(\la_2)=\mathcal{L}(\la_1,\la_2)$ where $\Gamma^X$ are actually the leptonic parts of the NP operators given in Eq.~(\ref{eq:operatorWC}). The spinors of particle $u$ and anti-particle $v$ are defined as 
\[u_{\frac{1}{2}}=\begin{pmatrix}
\sqrt{E_1 + m_{\lone}}\\
0\\
\sqrt{E_1 - m_{\lone}}\\0
\end{pmatrix},
u_{-\frac{1}{2}}=\begin{pmatrix}
0\\
\sqrt{E_1 + m_{\lone}}\\
0\\
-\sqrt{E_1 - m_{\lone}}
\end{pmatrix},
v_{\frac{1}{2}} = \begin{pmatrix}
\sqrt{E_2 - m_{\ltwo}}\\ 0\\-\sqrt{E_2 + m_{\ltwo}}\\0\end{pmatrix},
v_{-\frac{1}{2}} =\begin{pmatrix} 
0\\\sqrt{E_2 - m_{\ltwo}}\\ 0\\\sqrt{E_2 + m_{\ltwo}}\end{pmatrix}
\]
where the lepton energies are defined as $E_{1,2} = \sqrt{m^2_{1,2}+\lambda(q^2,m^2_1,m^2_2)/4q^2}$ and it gives $E_1+E_2 = \sqrt{q^2}$. The spinors are normalised as  $\bar u(\la_1) u(\la_2)   = \delta_{\la_1\la_2} 2m_{\lone}$ and 
 $\bar v(\la_1) v(\la_2)   = - \delta_{\la_1\la_2} 2m_{\ltwo}$.
 Using these, we find the following expressions of lepton helicity amplitudes
 \begin{eqnarray}
\mathcal{L}^{L,R}(1/2,1/2,0)&=& \left(\mm \bpl \pm \mpl \bmi\right)/2\\
\mathcal{L}^{L,R}(-1/2,-1/2,0)&=& (\mm \bpl \mp \mpl \bmi)/2\\
\mathcal{L}^{L,R}(1/2,1/2,1)&=& (\mpl \bmi \pm \mm \bpl  )/2\\
\mathcal{L}^{L,R}(-1/2,-1/2,1)&=&(\mpl \bmi \mp \mm \bpl  )/2 \\
\mathcal{L}^{L,R}(-1/2,1/2,1)&=& -\sqrt{q^2}(  \bmi  \pm \bpl )/\sqrt{2} \\
\mathcal{L}^{L,R}(1/2,-1/2,1)&=& -\sqrt{q^2}(  \bpl  \mp \bmi )/\sqrt{2} \\
\mathcal{L}^{S}(\pm 1/2, \pm 1/2)& =& \sqrt{q^2}\bpl \\
\mathcal{L}^{P}(\pm 1/2, \pm 1/2) &=& \mp \sqrt{q^2}\bmi
\end{eqnarray}
where $\beta_{\pm} = \sqrt{1-\frac{(m_{\ell_{1}} \pm m_{\ell_{2}})^2}{q^2}} $ and $m_{
\pm}=(m_1\pm m_2)$ as before. The lepton helicity amplitudes not mentioned above are all zero. Note that the scalar transitions $S$ and $P$ (also the timelike ones) are diagonal since $\la_\ell = \la_1 -\la_2 =0$.


\end{document}